\documentclass[aip, cp, amsmath, amssymb, reprint]{revtex4-2}

\usepackage{graphicx}
\usepackage{dcolumn}
\usepackage{bm}
\usepackage[utf8]{inputenc}
\usepackage[T1]{fontenc}
\usepackage{mathptmx}
\usepackage{hyperref}
\usepackage{float}
\usepackage{amsmath}
\usepackage{amssymb}
\hypersetup{
  setpagesize  = false,
  colorlinks   = true,    
  urlcolor     = blue,    
  linkcolor    = black,   
  citecolor    = black    
}
\begin{document}

\title{Exploring the Interplay Between Quantum Entanglement and Decoherence}

\author{Samuel A. Márquez González\vspace{1em}}

\affiliation{Department of Physics\looseness=-1, University of Maryland\looseness=-2, College Park\looseness=-3, MD\looseness=-4, 20742\looseness=-5, USA\vspace{1em} \textit{\centerline{Corresponding author: samarquz@terpmail.umd.edu\vspace{1em}}}}

\begin{abstract}
\hspace*{-0.1cm}Quantum entanglement manifests as a distinctive correlation between particles that transcends classical boundaries when their quantum states cannot be described independently. On the other hand, as quantum systems interact with their surroundings, decoherence emerges, leading to the gradual decay of quantum coherence and entanglement. In the case of entanglement, this is known as entanglement sudden death (ESD). Decoherence mechanisms are examined, focusing on how various environmental factors, such as thermal, electromagnetic, and collisional decoherence, influence the integrity of entangled states. The role of quantum noise, such as amplitude damping, phase damping, and depolarizing, is also analyzed. By integrating theoretical insights with experimental findings, this study highlights the delicate balance between maintaining entanglement and mitigating decoherence. The findings have significant implications for the development of quantum technologies, including quantum computing and quantum communication, where preserving entanglement is crucial for achieving robust and reliable performance.
\end{abstract}

\maketitle

\vspace{-2em}

\section{Introduction}

Quantum entanglement and decoherence are two fundamental concepts in the field of quantum mechanics that have significantly influenced our understanding of the physical world. Entanglement, first discussed by Albert Einstein, Boris Podolsky, and Nathan Rosen in 1935 and later formalized by Erwin Schrödinger, describes a phenomenon where the quantum states of two or more particles become intertwined, such that the state of one particle instantaneously affects the state of the other, regardless of the distance separating them. This "spooky action at a distance," as Einstein famously dubbed it, challenged the classical notions of locality and determinism, opening up new avenues for theoretical and experimental exploration.

States that can be expressed as a product of pure states are called separable states. Otherwise, they are considered entangled states \cite{Entanglement_and_Decoherence_Foundations_and_Modern_Trends}. Separability is defined by the ability to decompose a state into product states for pure states or into a convex sum of tensor products for mixed states \cite{Entanglement_and_Decoherence_Foundations_and_Modern_Trends}. Thus, entanglement can be described using the formalism of quantum states and tensor products. Consider two quantum systems, A and B, with respective states \(|\psi_A\rangle\) and \(|\psi_B\rangle\). The combined state of the system is given by the tensor product \(|\psi\rangle = |\psi_A\rangle \otimes |\psi_B\rangle\). An entangled state, however, cannot be factored into such a product of individual states. Thus, the dense matrix \(\rho\), which is a positive semidefinite operator on the space described by the tensor product, is called entangled when it cannot be written in the following form \cite{decteting_quantum_entanglement}:

\begin{equation}\label{eqn:equ1}
\rho = \sum_i p_i \left( |\psi_A^i\rangle \langle \psi_A^i| \otimes |\psi_B^i\rangle \langle \psi_B^i| \right), 
\quad p_i \geq 0, \; \sum_i p_i = 1
\end{equation}

In this form, \(\rho\) represents a separable state if it can be expressed as a sum of product states, with \(\rho_i\) being probabilities that sum to one. If such a decomposition is not possible, the state is entangled, reflecting the non-classical correlations between the subsystems \cite{decteting_quantum_entanglement}.

For instance, the Bell state \(|\Phi^+\rangle = \frac{1}{\sqrt{2}} (|00\rangle + |11\rangle)\) is a maximally entangled state of two qubits, where the measurement outcome of one qubit is perfectly correlated with the other, demonstrating non-local correlations \cite{Quantum_information_theory}. It is possible to achieve high-quality entanglement, one can employ methods like high-fidelity entanglement generation, quantum error correction codes (QECCs), and entanglement purification protocols (EPPs). Nevertheless, imperfections in physical devices can degrade the quality of the entanglement \cite{Advances_in_quantum_entanglement_purification}.

On the other hand, the quantum-to-classical transition problem was first articulated nearly fifty years ago by Zeh \cite{On_the_interpretation_of_measurement_in_quantum_theory}, leading to the development of the theory of quantum decoherence, also known as dynamical decoherence or environment-induced decoherence, which is the loss of quantum superposition. The insight is that realistic quantum systems are never completely isolated from their environment. When a quantum system interacts with its environment, it generally becomes rapidly and strongly entangled with numerous environmental degrees of freedom. This entanglement significantly affects what we can observe locally when measuring the system \cite{quantum_decoherence}. This interaction causes the superposition states to become entangled with the states of the environment, leading to the apparent collapse of the wave function and the emergence of classical behavior. Also, this interaction destroys the original entanglement of entangled quantum systems, which is known as entanglement sudden death (ESD) \cite{Decoherence-Induced_Sudden_Death_of_Entanglement_and_Bell_Nonlocality}. The interaction between a quantum system and its (macroscopic or mesoscopic) environment effectively acts as a measurement. Consequently, the environment's state is influenced by the system's state, meaning some information about the system's state is captured in the state of the environment (reservoir) \cite{Decoherence_and_dissipation_from_theory_of_continuous_measurements}. Decoherence thus bridges the gap between quantum mechanics and classical physics, providing a framework for understanding the quantum-to-classical transition.

One of the most notable aspects of the decoherence phenomenon is its remarkable efficiency, particularly in mesoscopic and macroscopic quantum systems. Furthermore, the multitude of uncontrollable environmental degrees of freedom generally renders the entanglement between the system and its environment practically irreversible. This effective irreversibility is a defining characteristic of the decoherence process \cite{Advances_in_quantum_entanglement_purification}. Decreasing decoherence in quantum systems is one of the principal goals of quantum information research, and this would help in quantum telecommunications and quantum computing.

Decoherence can be described using the density matrix formalism. The state of a quantum system is represented by a density operator \(\rho\) on a Hilbert space \(\mathcal{H}\). The system interacts with a single environmental degree of freedom at a time, such as a phonon, polaron, or gas particle. Assume this environmental "particle" is in a pure state \(\rho_e = |\psi_{\text{in}}\rangle \langle \psi_{\text{in}}|_e\), with \(|\psi_{\text{in}}\rangle_e \in \mathcal{H}_e\). The scattering operator \(\mathcal{S}_{\text{tot}}\) maps between the in- and out-asymptotes in the total Hilbert space, and for sufficiently short-ranged interaction potentials, we can identify these with the states before and after the collision. The equations before and after the collision are: \cite{Entanglement_and_Decoherence_Foundations_and_Modern_Trends}.

\begin{equation}\label{eqn:equ2}
\text{(before the collision)} \quad \rho_{\text{tot}} = S_{\text{tot}} \left[\rho \otimes |\psi_{\text{in}}\rangle \langle \psi_{\text{in}}|_e\right]
\end{equation}

\begin{equation}\label{eqn:equ3}
\text{(after the collision)} \quad \rho'_{\text{tot}} = S_{\text{tot}} \left[\rho \otimes |\psi_{\text{in}}\rangle \langle \psi_{\text{in}}|_e\right] S_{\text{tot}}^*
\end{equation}

Additionally, let's assume the interaction is non-invasive concerning a specific system property. This implies that there are several distinct system states, and the environment scattering off these states does not cause transitions within the system. Denoting the set of these mutually orthogonal system states by \(\{|n\rangle\} \in \mathcal{H}\), the non-invasiveness requirement means that \(\mathcal{S}_{\text{tot}}\) commutes with these states, i.e., it has the form \cite{Entanglement_and_Decoherence_Foundations_and_Modern_Trends}:

\begin{equation}\label{eqn:equ4}
S_{\text{tot}} = \sum_{\sigma_n} [|n\rangle \langle n| \otimes S_n]
\end{equation}

Consequently, substituting equation (4) into equation (3) results in \cite{Entanglement_and_Decoherence_Foundations_and_Modern_Trends}:

\begin{equation}\label{eqn:equ5}
\rho'_{\text{tot}} = \sum_{\sigma_{m,n}} \left[ \langle m|\rho|n \rangle \langle n| \otimes S_m |\psi_{\text{in}}\rangle \langle \psi_{\text{in}}|_e S^*_n \right] \equiv \sum_{\sigma_{m,n}} \left[ \rho_{m,n} |m\rangle \langle n| \otimes |\psi^{(m)}_{\text{out}}\rangle \langle \psi^{(n)}_{\text{out}}|_e \right]
\end{equation}

In the presence of decoherence, the off-diagonal elements of the density matrix, which represent quantum coherence, decay over time. For instance, consider a system initially in a superposition state involving two components. The corresponding density matrix includes terms that show the probabilities of each component and their quantum coherence. Due to interaction with the environment, the elements that represent the coherence between these components decay exponentially, resulting in a mixed state where the coherence is lost \cite{Quantum_Decoherence_and_Qubit_Devices}.

The interplay between entanglement and decoherence is of paramount importance in modern physics, particularly in the realms of quantum computing and quantum communication. Entanglement is a key resource for many quantum technologies, enabling phenomena such as quantum teleportation, superdense coding, and quantum cryptography. It is also integral to the operation of quantum computers, where entangled qubits provide exponential computational power over classical bits. However, the same interactions with the environment that lead to decoherence pose a significant challenge to maintaining entangled states, as they tend to disrupt the delicate quantum correlations essential for these technologies. Understanding the mechanisms of decoherence and developing strategies to mitigate its effects are crucial for the advancement of quantum technologies. Researchers are exploring various techniques, such as error correction codes \cite{Quantum_Computing_and_Error_Correction}, decoherence-free subspaces \cite{Decoherence-Free_Subspaces_and_Subsystems}, and dynamical decoupling \cite{Optimized_dynamical_decoupling_in_a_model_quantum_memory}, to preserve entanglement and ensure the reliable operation of quantum systems. The ability to control and harness entanglement while counteracting decoherence holds the key to unlocking the full potential of quantum computing, secure communication, and other emerging quantum applications. Thus, this paper will explore how the technical effects of decoherence affect entanglement in quantum systems.

\section{Environmental Interactions}

Quantum systems are highly sensitive to their surrounding environment. Any realistic quantum system will inevitably be influenced by its environment, making the dynamic characteristics of open quantum systems especially significant from a practical standpoint \cite{Evolution_of_coherence_and_non-classicality_under_global}. Various environmental factors can induce decoherence, which disrupts the delicate entanglement between quantum particles. Understanding how different environmental elements affect quantum systems is crucial for developing strategies to mitigate these effects and preserve entanglement.

Environmental factors are taken into serious consideration when researchers work on quantum technologies. Environmental perturbations have been observed to introduce a new aspect to the entanglement dynamics between two qubits. As these initially entangled qubits interact and evolve over time, they undergo repeated cycles of disentanglement and re-entanglement, resulting in alternating phases of high and low entanglement. Ultimately, after prolonged periods, this process leads to entanglement sudden death (ESD) \cite{Bright_and_dark_periods_in_the_entanglement_dynamics}.

\subsection{Thermal Decoherence}

Thermal decoherence is a critical factor in the degradation of quantum coherence and entanglement in quantum systems. When a quantum system interacts with a thermal environment, the thermal fluctuations induce interactions with phonons \cite{Phonon_heat_transfer_across_a_vacuum_through_quantum_fluctuations}, leading to decoherence. This process can significantly impact the entanglement between quantum particles, reducing their quantum correlations and transitioning the system toward classical behavior. In the context of entanglement, thermal decoherence can lead to entanglement sudden death (ESD), where the entangled state abruptly loses its coherence \cite{Entanglement_evolution,Thermodynamics_of_decoherence}.

\begin{figure}[H]
    \centering
    \includegraphics[width=0.5\linewidth]{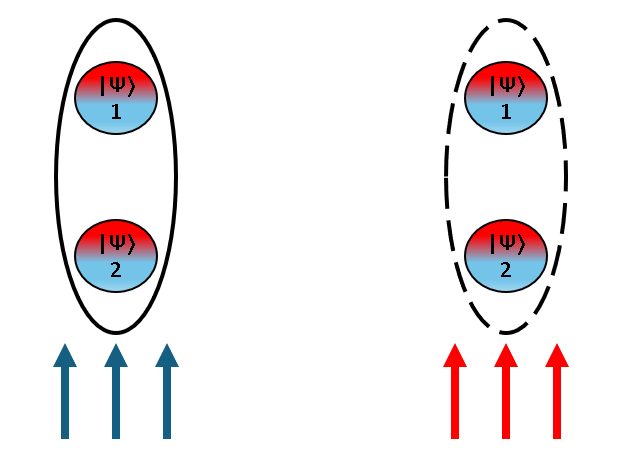}
    \caption{Interaction of entangled states under varying thermal conditions. The interaction on the left represents entanglement at low temperatures, whereas the interaction on the right corresponds to high temperatures. It is evident that entanglement is better preserved at lower thermal decoherence (low temperature).}
    \label{fig:enter-label}
\end{figure}

The formalism of superstatistics provides a valuable framework for understanding entanglement in systems subject to temperature fluctuations. Superstatistics involves a superposition of different statistics, each corresponding to varying environmental conditions over time. This approach is particularly useful for describing systems in slowly fluctuating environments with weak interactions between the system and its surroundings \cite{Quantum_entanglement_and_temperature_fluctuations}.

The relationship between temperature and decoherence is quantitatively described by the master equation for finite temperatures. The master equation models the time evolution of the system's density matrix, incorporating the effects of thermal excitations. At zero temperature, the system can retain a non-zero steady-state concurrence despite decoherence. However, as the temperature increases, the steady-state concurrence decreases and eventually vanishes at high temperatures. This behavior is captured by the formula for concurrence at finite temperatures \cite{Entanglement_evolution}:

\begin{equation} 
C(\bar{\omega}, \Delta, \bar{n}) = 2 \frac{\sqrt{\Delta^{2}(4\bar{\omega}^{2} + (1 + 2\bar{n})^{2})}}{(1 + 2\bar{n})(4\bar{\omega}^{2} + (1 + 2\bar{n})^{2})} - \frac{1}{2} + \frac{(4\bar{\omega}^{2} + (1 + 2\bar{n})^{2})}{2(1 + 2\bar{n})^{2}(4\bar{\omega}^{2} + (1 + 2\bar{n})^{2})}
\end{equation} 

In the given equation, \( C(\bar{\omega}, \Delta, \bar{n}) \) represents the concurrence, which depends on the normalized energy difference \( \bar{\omega} \), the interaction anisotropy \( \Delta \), and the average thermal excitation \( \bar{n} \). The terms in the equation encapsulate the contributions of these factors: the first term accounts for the impact of the energy difference and thermal excitations on the entanglement, the second term represents a baseline entanglement reduction due to decoherence, and the third term considers the enhancement or suppression of entanglement by the thermal environment, adjusted for the interaction anisotropy.

The impact of thermal decoherence on entanglement is not limited to low-dimensional systems. Recent studies have shown that even in higher-dimensional systems, such as spin chains, quantum coherence can persist at temperatures significantly higher than the interaction energy scale. For instance, neutron scattering experiments on YbAlO$_3$ spin chains have revealed the persistence of quantum coherence in spinon excitations at temperatures where thermal energy exceeds the interaction energy by more than an order of magnitude. This remarkable resilience of quantum coherence in the face of thermal fluctuations challenges the conventional understanding of thermal decoherence and has profound implications for the development of quantum technologies \cite{High_temperature_coherence}.

Moreover, the thermodynamics of decoherence provides insights into the energetic costs associated with maintaining quantum coherence. In purely decoherent processes, where there is no direct energy exchange between the system and its environment, heat dissipation occurs due to the interactions with the thermal bath. This dissipation can be quantified and analyzed using the principles of non-equilibrium thermodynamics. For example, the heat generated by decoherence in a two-level system interacting with a thermal bath can be described by the change in the von Neumann entropy and the average heat absorbed by the bath, emphasizing the energetic implications of maintaining quantum coherence in thermal environments \cite{Thermodynamics_of_decoherence}.

\subsection{Electromagnetic Decoherence}

Quantum entanglement is significantly influenced by the interaction with electromagnetic fields. These interactions can enhance or degrade the entanglement depending on the characteristics of the field and the system's parameters. In particular, the exact solution of the Schrödinger equation for a quantum harmonic oscillator interacting with a quantized electromagnetic field reveals the conditions under which significant entanglement can be achieved \cite{Quantum_entanglement_of_a_harmonic_oscillator_with_an_electromagnetic_field}. This interaction is crucial for understanding and harnessing quantum entanglement in practical applications.

Electromagnetic fields, both static and dynamic, introduce external perturbations that can either facilitate or disrupt the coherent superposition of quantum states. For instance, static fields can shift energy levels and modify the potential landscape, thus altering the dynamics of entanglement. Dynamic fields, on the other hand, can induce transitions between quantum states, affecting the entanglement properties. The study of such interactions is crucial for developing robust quantum systems that can maintain high levels of entanglement over time \cite{Quantum_entanglement_of_a_harmonic_oscillator_with_an_electromagnetic_field}.

The Schrödinger equation for the system that has the interaction of a harmonic oscillator of mass m via an electric charge e \cite{Quantum_entanglement_of_a_harmonic_oscillator_with_an_electromagnetic_field}:

\begin{equation} 
i \hbar \frac{\partial \Psi}{\partial t} = \left[ \frac{1}{2m} \left( -i \hbar \frac{\partial}{\partial \mathbf{r}} + \frac{e}{c} \mathbf{\hat{A}} \right)^2 + \hat{H}_f + \frac{m \omega_c^2 (x^2 + \delta(y^2 + z^2))}{2} \right] \Psi
\end{equation}

Where \(\delta = 0\) corresponds to one type of interaction, and \(\delta = 1\) to another type. \(i\) represents the imaginary unit, \(\omega_c\) is the frequency of the oscillator, and \(\hat{H}_f\) denotes the Hamiltonian of the electromagnetic field, which is formulated as follows\cite{Quantum_entanglement_of_a_harmonic_oscillator_with_an_electromagnetic_field}:

\begin{equation}
\hat{H}_f = \sum_{k, u} \hbar \omega \left( \hat{a}_{k, u}^\dagger \hat{a}_{k, u} + \frac{1}{2} \right)
\end{equation}

Here, \(\hat{a}_{k, u}^\dagger\) and \(\hat{a}_{k, u}\) are the creation and annihilation operators, respectively, for photons characterized by the wave vector \(k\) and polarization \(u\). The vector potential \(\mathbf{\hat{A}}\) of the electromagnetic field is expressed as\cite{Quantum_entanglement_of_a_harmonic_oscillator_with_an_electromagnetic_field}:

\begin{equation}
\mathbf{\hat{A}} = \sum_{k, u} \sqrt{\frac{2 \pi c^2 \hbar}{\omega V_f}} \left( \mathbf{u} \exp(i(\omega t - \mathbf{kr})) \hat{a}_{k, u} + \mathbf{u}^* \exp(-i(\omega t - \mathbf{kr})) \hat{a}_{k, u}^\dagger \right)
\end{equation}

\begin{figure}[H]
    \centering
    \includegraphics[width=0.63\linewidth]{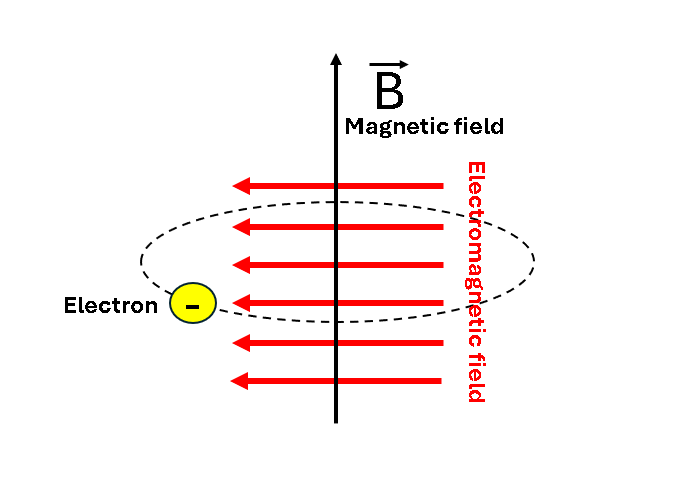}
    \caption{Interaction between a magnetic field with an electron and an electromagnetic field (quantum harmonic oscillator).
}
    \label{fig:enter-label}
\end{figure}

Thus, quantum entanglement is maximized when the oscillator's frequency \( \omega_c \) closely matches the frequency of the electromagnetic field \( \omega \), with a small difference \( \Delta \omega \) on the order of the parameter \( \beta \), where \( \beta = \sqrt{\frac{\pi \omega V}{4}} \) and \( V \) is the quantization volume. When these frequencies are nearly identical, the system reaches significant entanglement over a longer time \( t \sim \frac{1}{\beta} \), while large mismatches in frequency result in weaker entanglement \cite{Quantum_entanglement_of_a_harmonic_oscillator_with_an_electromagnetic_field}.

In the context of non-inertial reference frames, the study by Ling et al. provides valuable insights into how quantum entanglement is influenced by acceleration and the Unruh effect. The research highlights that the photon helicity entangled state remains invariant in terms of logarithmic negativity and mutual information, even under acceleration. This invariance contrasts with the degradation observed in particle number entangled states, showcasing the unique robustness of helicity entangled states against acceleration-induced perturbations \cite{Ling2007}.

Electromagnetic noise introduces random fluctuations that disrupt the coherent evolution of quantum states. Understanding these interactions helps in devising strategies to mitigate their adverse effects and preserve entanglement\cite{Quantum_entanglement_of_a_harmonic_oscillator_with_an_electromagnetic_field}. Recent studies have shown that electromagnetic fields can be used to control and manipulate quantum entanglement. By carefully tuning the frequency and amplitude of the applied field, it is possible to enhance the entanglement between particles. This technique is particularly useful in quantum computing, where controlled entanglement is necessary for performing quantum gate operations and error correction protocols. The ability to manipulate entanglement with electromagnetic fields opens new avenues for developing advanced quantum technologies \cite{Quantum_entanglement_of_a_harmonic_oscillator_with_an_electromagnetic_field}.

The interplay between quantum entanglement and electromagnetic fields also has implications for quantum communication and cryptography. Entangled particles can be used to establish secure communication channels, where the presence of eavesdropping can be detected through disturbances in the entanglement. Electromagnetic fields play a crucial role in the generation and transmission of entangled states over long distances, making them essential for practical implementations of quantum networks\cite{Quantum_entanglement_of_a_harmonic_oscillator_with_an_electromagnetic_field}.

The interaction between quantum systems and electromagnetic fields is a double-edged sword. While it can lead to decoherence and loss of entanglement, it also offers tools for controlling and enhancing entanglement. The exact solutions of the Schrödinger equation provide valuable insights into the conditions required for achieving high levels of entanglement, paving the way for practical applications in quantum computing, communication, and beyond. Future research will continue to explore these interactions, aiming to optimize the balance between maintaining entanglement and mitigating decoherence\cite{Quantum_entanglement_of_a_harmonic_oscillator_with_an_electromagnetic_field}. Also, The findings from Ling et al.'s research further underscore the importance of understanding these interactions, especially in non-inertial frames, to optimize the balance between maintaining entanglement and mitigating decoherence \cite{Ling2007}.

\subsection{Collisional Decoherence}

Collisional decoherence occurs when environmental particles scatter off a massive free quantum particle. This scattering process allows the environmental particles to gain information about the path taken by the central particle \cite{quantum_decoherence}. When dealing with entangled states, this process becomes particularly intricate. For an entangled state, this process can be understood through the master equation, which describes the time evolution of the system's density matrix. The master equation for collisional decoherence, assuming a massive central particle compared to the environmental particles, can be written as \cite{quantum_decoherence}:

\begin{equation}
\frac{\partial \rho_S(x, x', t)}{\partial t} = -F(x - x')\rho_S(x, x', t),
\end{equation}
where \(F(x - x')\) is the decoherence factor. This factor represents the characteristic decoherence rate at which coherence between two positions \(x\) and \(x'\) becomes locally unobservable. It is given by \cite{quantum_decoherence}:

\begin{equation}
F(x - x') = \int \limits_{0}^{\infty} dq \rho(q) v(q) \int dn \, dn' \, \frac{1}{4\pi} \left(1 - e^{iq(n-n')\cdot (x-x')/\hbar}\right) |f(qn, qn')|^2,
\end{equation}
where \(\rho(q)\) is the number density of incoming environmental particles, \(v(q)\) is their speed, and \(f(qn, qn')\) is the differential cross-section for the scattering of an environmental particle.

\begin{figure}[H]
    \centering
    \includegraphics[width=0.35\linewidth]{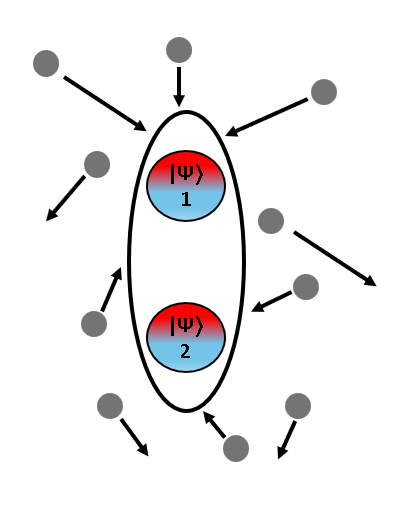}
    \caption{Interaction between an entangled state system and the particles in the environment, causing collisional decoherence.}
    \label{fig:enter-label}
\end{figure}

Experimental studies on collisional decoherence have shown excellent agreement between theoretical predictions and observed data. For instance, experiments with fullerene molecules in a Talbot-Lau interferometer have demonstrated the decoherence effect due to collisions with background gas molecules \cite{quantum_decoherence}. The interference fringes in these experiments showed a decrease in visibility with increasing pressure of the background gas, confirming the theoretical models of collisional decoherence based on the quantum linear Boltzmann equation \cite{Limitations_of_squeezing}.

Further advancements in the theoretical framework have incorporated the effects of orientational degrees of freedom, particularly in systems involving anisotropic molecules. These extended models account for both spatial and orientational decoherence, providing a more comprehensive understanding of collisional decoherence in complex systems \cite{quantum_decoherence}. Such developments highlight the ongoing interplay between theory and experiment in the study of quantum decoherence.

The impact of collisional decoherence on entanglement is profound, as the interactions with environmental particles lead to the rapid loss of quantum coherence, crucial for maintaining entangled states. Studies have shown that even weak collisional interactions can result in entanglement sudden death (ESD), where the entangled state abruptly loses its quantum correlations \cite{Limitations_of_squeezing}. This phenomenon has significant implications for quantum information processing and quantum computing, as it limits the practical use of entangled states in real-world environments \cite{Limitations_of_squeezing}. To mitigate these effects, various strategies such as error correction codes and entanglement purification protocols are being explored \cite{quantum_decoherence}.

\section{Quantum Noise}

Quantum noise introduces various disturbances that can degrade the quality of quantum entanglement in a system. This type of noise includes amplitude damping, phase damping, and depolarizing. Each of these disturbances leads to the loss of coherence and the transition of quantum states from pure to mixed states. Amplitude damping involves the loss of energy from the system to the environment, which reduces the amplitude of the quantum state. Phase damping, or dephasing, causes the decay of the off-diagonal elements of the density matrix, resulting in the loss of coherence without energy exchange. Depolarizing noise represents a generalized form of noise where the quantum state is randomly replaced by the maximally mixed state with a certain probability.

The presence of quantum noise significantly impacts the integrity of entangled states, making it challenging to maintain high-quality entanglement over time. Understanding the mechanisms and effects of these types of noise is essential for developing strategies to mitigate their impact. Effective mitigation techniques can help preserve entanglement and ensure the reliable operation of quantum systems, which is crucial for advancing quantum computing and other quantum technologies.

\subsection{Amplitude Damping}

Amplitude damping is a significant source of quantum noise, characterized by the gradual loss of energy from a quantum system to its environment. This type of decoherence typically occurs in systems where excited states decay to lower energy states, such as in atomic and photonic systems. The process can be described by the amplitude damping channel, which models the interaction between the quantum system and its surroundings, often resulting in a spontaneous emission of photons \cite{Quenched_decoherence_in_qubit_dynamics_due_to_strong_amplitude-damping_noise}. This loss of energy alters the probabilities associated with the quantum states, leading to a reduction in the amplitude of the superposition components. Consequently, the quantum information encoded in the system becomes less reliable over time. The mathematical formalism of amplitude damping involves a Kraus operator representation, where the evolution of the density matrix is governed by specific operators that capture the probabilistic nature of energy decay \cite{Quenched_decoherence_in_qubit_dynamics_due_to_strong_amplitude-damping_noise}. Understanding and mitigating amplitude damping is crucial for the development of robust quantum technologies, as it directly impacts the fidelity of quantum states in computing and communication applications \cite{Quenched_decoherence_in_qubit_dynamics_due_to_strong_amplitude-damping_noise}.

In systems subject to amplitude damping noise, the Kraus operators \( E_0 \) and \( E_1 \) describe the evolution of the quantum state. These operators are given by \cite{Preskill1998}:

\begin{equation}
E_0 = \begin{pmatrix}
1 & 0 \\
0 & \sqrt{1 - \gamma}
\end{pmatrix}, \quad E_1 = \begin{pmatrix}
0 & \sqrt{\gamma} \\
0 & 0
\end{pmatrix},
\end{equation}

where \( \gamma \) represents the probability of the system transitioning to the ground state. The density matrix \(\rho\) evolves according to \cite{Preskill1998}:

\begin{equation}
\rho' = E_0 \rho E_0^\dagger + E_1 \rho E_1^\dagger.
\end{equation}

This formalism effectively captures the dynamics of energy dissipation in the system, leading to a gradual reduction in coherence. The work by Shin-Tza Wu demonstrates that strong coupling between a qubit and its environment can result in quenching of decoherence, thereby preserving quantum coherence even in the presence of strong amplitude-damping noise \cite{Quenched_decoherence_in_qubit_dynamics_due_to_strong_amplitude-damping_noise}. This phenomenon arises due to the large feedback from the environment to the qubit, which keeps the qubit's coherence intact after an initial period of decay. Mathematically, this can be represented as\cite{Quenched_decoherence_in_qubit_dynamics_due_to_strong_amplitude-damping_noise}:

\begin{equation}    
d\rho_s(t) = -i[S(t)[\sigma_+\sigma_-, \rho_s(t)]] + \gamma(t)[2\sigma_-\rho_s(t)\sigma_+ - \sigma_+\sigma_-\rho_s(t) - \rho_s(t)\sigma_+\sigma_-],
\end{equation}

where \(S(t)\) and \(\gamma(t)\) are time-dependent functions representing the frequency shift and decay rate, respectively.

Quantum entanglement, a crucial resource in quantum information processing, is particularly vulnerable to decoherence. High-dimensional systems, such as qutrits (three-level systems), are being explored for their potential to offer greater robustness against noise compared to qubits. The study by Xiao et al. investigates the protection of three-dimensional entanglement from correlated amplitude damping (CAD) noise using weak measurement (WM) and environment-assisted measurement (EAM) combined with quantum measurement reversal (QMR) \cite{Protecting_three-dimensional_entanglement_from_correlated_amplitude_damping_channel}. The Kraus operators for a qutrit under CAD noise can be expressed as:

\begin{equation}
E_0 = \begin{pmatrix}
1 & 0 & 0 \\
0 & \sqrt{1 - d_1} & 0 \\
0 & 0 & \sqrt{1 - d_2}
\end{pmatrix}, \quad E_1 = \begin{pmatrix}
0 & \sqrt{d_1} & 0 \\
0 & 0 & 0 \\
0 & 0 & 0
\end{pmatrix}, \quad E_2 = \begin{pmatrix}
0 & 0 & \sqrt{d_2} \\
0 & 0 & 0 \\
0 & 0 & 0
\end{pmatrix},
\end{equation}

where \(d_1\) and \(d_2\) are the probabilities of transitions to lower energy states. The density matrix \(\rho\) for the qutrit evolves as:

\begin{equation}
\rho' = E_0 \rho E_0^\dagger + E_1 \rho E_1^\dagger + E_2 \rho E_2^\dagger.
\end{equation}

The effectiveness of these methods in protecting entanglement is quantified using measures such as negativity and success probability. For instance, in the EAM+QMR scheme, the negativity \(N\) of the entangled state can be significantly restored, as shown in the equation:

\begin{equation} 
N = \frac{\| \rho^{T_1} \| - 1}{2},
\end{equation}

where \(\rho^{T_1}\) is the partial transpose of \(\rho\) and \(\|\cdot\|\) denotes the trace norm. The success probability of the entanglement recovery process is also a critical parameter, reflecting the likelihood of successfully implementing the correction without losing coherence.

The studies highlight the importance of understanding and mitigating amplitude damping in quantum systems. By employing strategies such as strong coupling, WM, and EAM, it is possible to preserve quantum coherence and entanglement, thereby enhancing the performance and reliability of quantum information processing technologies \cite{Quenched_decoherence_in_qubit_dynamics_due_to_strong_amplitude-damping_noise, Protecting_three-dimensional_entanglement_from_correlated_amplitude_damping_channel}.

\subsection{Phase Damping}

Phase damping noise, also known as phase decoherence, is a type of quantum noise that affects the phase of a quantum state without altering its energy. This type of decoherence is particularly insidious because it destroys the quantum coherence of superpositions, effectively transforming pure states into mixed states. Unlike amplitude damping, which involves energy loss to the environment, phase damping solely impacts the off-diagonal elements of the density matrix, representing the coherence between different states. This process leads to the decay of these off-diagonal elements, causing a loss of quantum coherence and affecting the overall entanglement of the system \cite{obada2015effects}.

The physical mechanism behind phase damping can be understood through interactions with the environment that randomize the phase relationships between components of the quantum state. For instance, in a two-level system, the relative phase between the two levels can become randomized due to fluctuations in the local environment, such as variations in electromagnetic fields or collisions with other particles. These environmental interactions cause dephasing, which results in the gradual decay of coherence. This loss of coherence is a significant challenge for quantum information processing, where maintaining precise phase relationships between qubits is crucial for operations like quantum gates and error correction \cite{ramzan2011decoherence}.

Phase damping significantly affects entangled states. Studies have shown that phase damping can lead to entanglement sudden death (ESD), where entanglement between particles vanishes abruptly and completely. For example, in a two-qubit system interacting with a phase-damped cavity, the entanglement can exhibit a phenomenon of death and rebirth depending on the phase damping parameter \cite{obada2015effects}. As the phase parameter increases, this phenomenon disappears, leading to a complete loss of entanglement over time. The mathematical representation of phase damping can be described using the Kraus operators \cite{Preskill1998}:

\begin{equation} 
E_0 = \begin{pmatrix}
1 & 0 \\
0 & \sqrt{1-\lambda}
\end{pmatrix}, \quad E_1 = \begin{pmatrix}
0 & 0 \\
0 & \sqrt{\lambda}
\end{pmatrix},
\end{equation} 

This formalism helps in quantifying the impact of phase damping on entangled states and provides insights into how this type of noise can be mitigated \cite{ramzan2012decoherence}.

To combat phase damping, various strategies have been proposed and implemented in quantum technologies. Techniques such as dynamical decoupling \cite{Optimized_dynamical_decoupling_in_a_model_quantum_memory}, where sequences of control pulses are applied to refocus the phase coherence, have shown promise in reducing the effects of phase damping. Additionally, decoherence-free subspaces, which exploit symmetries in the system to create states that are inherently immune to certain types of noise, offer another avenue for protecting quantum information. Understanding and mitigating phase damping is essential for the development of robust quantum computing and communication systems, as maintaining coherence is crucial for the performance and reliability of these technologies \cite{obada2015effects, ramzan2011decoherence, ramzan2012decoherence}. By continuing to explore and refine these strategies, researchers aim to enhance the stability and resilience of quantum systems in the face of environmental noise.

\subsection{Depolarizing}

Depolarizing noise is a type of quantum noise that affects the state of a quantum system by transforming it into a maximally mixed state with a certain probability. This process effectively randomizes the state of the quantum system, leading to a loss of coherence and entanglement. Depolarizing noise is often modeled as an interaction where, with a given probability, the quantum state is replaced by a completely mixed state, representing total uncertainty about the system's state. This type of noise is particularly detrimental to quantum information processing as it not only affects the phase and amplitude of the quantum state but also reduces the overall purity of the system \cite{Bavontaweepanya2018, Yashodamma2014}.

The depolarizing channel can be mathematically described using Kraus operators, which are essential in the operator-sum representation of quantum operations. The Kraus operators for the single-qubit depolarizing channel are given by \cite{Entanglement-Assisted_Communication_Capacity_2022}:

\begin{equation}
    E_1 = \sqrt{1 - p} \, I, \quad E_2 = \sqrt{\frac{p}{3}} \, X, \quad E_3 = \sqrt{\frac{p}{3}} \, Y, \quad E_4 = \sqrt{\frac{p}{3}} \, Z
\end{equation}

where \( I \) is the identity operator, and \( X, Y, Z \) are the Pauli matrices. The parameter \( p \) represents the depolarizing probability, and it ranges from 0 (no depolarization) to 1 (complete depolarization). The action of the depolarizing channel on a density matrix \( \rho \) is expressed as \cite{Entanglement-Assisted_Communication_Capacity_2022}:

\begin{equation}
    \rho' = \sum_{i=1}^{4} E_i \rho E_i^\dagger
\end{equation}

This equation shows that the original quantum state \( \rho \) is subjected to various random errors, resulting in a mixed state that approaches the maximally mixed state as \( p \) increases \cite{Bavontaweepanya2018, Entanglement_and_Decoherence_Foundations_and_Modern_Trends, Yashodamma2014}.

Depolarizing noise has a profound effect on entangled states. When entangled photons or qubits pass through a depolarizing channel, their entanglement can be significantly degraded. For example, the concurrence, a measure of entanglement, is reduced to zero in the presence of high levels of depolarizing noise. This degradation is due to the loss of quantum correlations between the entangled particles. Experimental studies have shown that even partial depolarizing noise can lead to a substantial reduction in entanglement quality, making it challenging to maintain high-fidelity quantum communication and computation \cite{Bavontaweepanya2018, Entanglement_and_Decoherence_Foundations_and_Modern_Trends, Yashodamma2014}.

To mitigate the effects of depolarizing noise, various strategies have been proposed. One such approach is the use of quantum error correction codes, which can protect quantum information by encoding it in a way that allows for the detection and correction of errors. Another method involves using entanglement purification protocols, which aim to distill high-quality entangled states from a larger set of noisy entangled states. Additionally, techniques like dynamical decoupling and decoherence-free subspaces have been explored to preserve entanglement in the presence of depolarizing noise \cite{Optimized_dynamical_decoupling_in_a_model_quantum_memory, Decoherence-Free_Subspaces_and_Subsystems}. These methods, while still under active research, hold promise for enhancing the robustness of quantum systems against the detrimental effects of depolarizing noise \cite{Bavontaweepanya2018, Entanglement_and_Decoherence_Foundations_and_Modern_Trends, Yashodamma2014}.

\section{Conclusion}

The exploration of the intricate relationship between quantum entanglement and decoherence underscores the delicate balance required to harness the potential of quantum systems. This study highlights the profound impact of various environmental factors, such as thermal fluctuations, electromagnetic fields, and collisional interactions, on the integrity of entangled states. Additionally, the analysis of quantum noise types, including amplitude damping, phase damping, and depolarizing noise, reveals the significant challenges posed by these disturbances to maintaining high-quality entanglement. Understanding these interactions is essential for advancing quantum technologies, where the preservation of entanglement is crucial for robust performance.

The insights gained from this research offer valuable guidance for developing strategies to mitigate the effects of decoherence. Techniques such as error correction codes, entanglement purification protocols, and the use of decoherence-free subspaces are pivotal in protecting entangled states from environmental disturbances. The ability to control and enhance entanglement in the presence of decoherence opens new avenues for practical applications in quantum computing, secure communication, and other emerging quantum technologies. As researchers continue to refine these methods, the reliability and efficiency of quantum systems will improve, paving the way for more advanced quantum applications.

Future research should focus on exploring different approaches to further understand and control the interplay between quantum entanglement and decoherence. Investigating the impact of more complex environmental interactions and developing new mitigation strategies will be key to overcoming the current limitations faced by quantum technologies. By mastering the delicate balance between entanglement preservation and decoherence mitigation, the full potential of the quantum world can be unlocked, leading to groundbreaking advancements in quantum information processing and beyond.

\bibliography{referencias}

\end{document}